\documentclass{elsart}
\usepackage{natbib}
\usepackage{epsfig}
\begin{document}
\begin{frontmatter}
\title{Hybrid Pixel Detector Development \\ for the Linear Collider
Vertex Tracker}
\author{M.~Battaglia\thanksref{X}}
\address{CERN, CH-1211 Geneva 23, Switzerland}
\author{S.~Borghi, R.~Campagnolo}
\address{Univ. degli Studi di Milano, Dip. di Fisica, Via Celoria 16,
Milano, Italy}
\author{M.~Caccia \thanksref{X}}
\thanks[X]{corresponding authors;\\ email: Marco.Battaglia@cern.ch and 
Massimo.Caccia@mi.infn.it}
\address{Univ. degli Studi dell'Insubria, Dip. di Scienze, Via Valleggio
11, Como, Italy}
\author{W.~Kucewicz}
\address{Univ. of Mining and Metallurgy, Dept. of Electronics, \\
al. Mickiewicza 30, Krakow, Poland}
\author{P.~Jalocha, H.~Palka, A.~Zalewska}
\address{High Energy Physics Lab., Institute of Nuclear Physics,\\
ul. Kawiory 26a, Krakow, Poland}
%
\begin{abstract}
In order to fully exploit the physics potential of the future high energy
$e^+e^-$ linear collider, a Vertex Tracker able to provide particle track 
extrapolation with very high resolution is needed. Hybrid Si pixel sensors are
an attractive technology due to their fast read-out capabilities and radiation 
hardness. A novel pixel detector layout with interleaved cells has been 
developed to improve the single point resolution. Results of the 
characterisation of the first processed prototypes by electrostatic 
measurements and charge collection studies are discussed.

\end{abstract}
\end{frontmatter}
\section{Introduction}

\vspace{-0.5cm} 

The physics programme at future high energy linear colliders, designed to 
deliver $e^+e^-$ collisions at centre-of mass energies $\sqrt{s}$ = 0.3 - 
3~TeV with luminosities in excess to $10^{34}$~cm$^{-2}$~s$^{-1}$, largely 
relies on the ability to identify the flavour of final state fermions with 
high efficiency and purity~\cite{challenges}.
This task must be accomplished in a challenging environment due to the large
track density from highly collimated hadronic jets and the accelerator induced
backgrounds. The estimated track density, $\ge$ 1 hit mm$^{-2}$, requires
the adoption of pixel detector technology.
The target performance for the resolution on the track impact
parameter, defined as the distance of closest approach to the $e^+e^-$ 
collision point, has been obtained from detailed studies of jet tagging 
performances in a variety of physics reactions. These results indicate that
a resolution of 5~$\mu$m $\oplus$ 15~$\mu$m / $p_t$~(GeV/c), or better, has to 
be achieved. Several conceptual designs have been proposed so far, relying on 
three alternative sensor technologies: Charge Coupled Devices 
(CCD)~\cite{ccd}, Hybrid Pixels~\cite{vertex99} and Monolithic 
CMOS sensors~\cite{cmos}.
This paper summarises recent results on the development of hybrid pixel
sensors for the future linear collider applications.

\vspace{-0.25cm} 

\section{Hybrid Pixel Sensor Design}

\vspace{-0.5cm} 

Hybrid pixel sensors, pioneered in fixed target and LEP experiments and further
developed for their application in the LHC Vertex Detectors, have several 
features of interest for application in the linear collider Vertex Tracker. 
Fast time-stamping capabilities and sparse data scan read-out allow to minimise
the effect of beam backgrounds, while their radiation hardness ensures 
a significant safety margin compared to the anticipated 
neutron fluence in the linear collider interaction region.
At present, their main limitation comes from the achievable single point 
resolution. A resolution $< 10~\mu$m, required to match the needed
impact parameter accuracy, can be obtained by sampling the diffusion of the 
carriers generated along the particle path and adopting an analog read-out to 
interpolate the signals of neighbouring cells. Since the charge diffusion 
r.m.s. in 300~$\mu$m thick silicon is $\simeq$~8~$\mu$m, its efficient sampling
requires a pixel pitch below 50~$\mu$m. As the most advanced read-out 
electronics have a cell dimension of 50 $\times$ 300~$\mu$m$^2$, this 
represents the present limit to the pixel pitch. Future developments in 
deep sub-micron VLSI~\cite{vlsi} may help overcoming this limit. However it is 
interesting to independently explore sensor designs improving the pixel sensor 
performances in terms of single point resolution.

The proposed pixel detector design exploits a layout, already successfully
adopted in Silicon microstrip detectors, where only one-out-of-$n$ implants 
is read-out. In such a 
configuration, charge carriers generated underneath one of the interleaved
pixel cells induce a signal on the capacitively coupled read-out pixels. The 
ratio of the signal amplitudes at the left- and right-hand side of the 
read-out pixels is proportional to the distance of the position of crossing 
of the particle from the read-out node. In this design the sampling of the 
charge
carrier distribution is achieved by the implant pitch and the analog cell size
has to fit the wider read-out pitch. The ratio between the inter-pixel, 
$C_{ip}$, and the pixel-to-backplane, $C_{bp}$, capacitances defines the signal
amplitude reduction by charge loss to the backplane. This signal reduction and
the two track separation performances limit the number $n$ of interleaved 
pixels.   

\vspace{-0.25cm} 

\section{Hybrid Pixel Sensor Tests}

\vspace{-0.5cm} 

Prototypes of detectors with interleaved pixels have been designed and 
manufactured~\cite{sitges}. Thirty-six test structures have been fit on a 
4$^{''}$ wafer, consisting
of detectors with 0 to 3 interleaved pixels defining a VLSI cell size of
either 200 $\times$ 200~$\mu$m$^2$ or 300 $\times$ 300~$\mu$m$^2$. The
structures have been characterised by electrostatic tests. About half of the
tested structures exhibit a leakage current of 10~nA/cm$^2$, or less, at full 
depletion and a breakdown point well beyond the full depletion voltage. 
The inter-pixel and backplane capacitances, crucial to determine the 
feasibility of the proposed sensor layout, have been directly measured on 
different structures and the results compared to a numerical estimate 
obtained by solving the Laplace equation inside a 5$\times$5 pixel matrix 
with a finite element analysis package~\cite{opera}. 
The results have been found to be in fairly good agreement. The computed values
for the single pixel capacitance are summarised in Table~1.

\begin{table}[h!]
\begin{center}
\caption{\sl Computed $C_{ip}$ and $C_{bp}$ values for different detector
structures.}
\begin{tabular}{l c c c c}
\hline
   & Det~1 & Det~2 & Det~3 & Det~4 \\
\hline \hline
Implant width ($\mu$m) & 100 & 60 & 50 & 34 \\
Implant pitch ($\mu$m) & 150 & 100 & 75 & 50 \\ \hline
$C_{ip}$ total ($fF$) & 22.3$\pm$1.1 & 12.7$\pm$0.6 & 11.8$\pm$0.6 & 
~8.6$\pm$0.4\\
$C_{ip}$ to nearest ($fF$) & ~3.7$\pm$0.2 & ~1.9$\pm$0.1 & ~1.8$\pm$0.1 & 
~1.3$\pm$0.1\\
$C_{bp}$ ($fF$) & ~7.3$\pm$1.1 & ~3.2$\pm$0.7 & ~1.9$\pm$0.7 & 
~0.8$\pm$0.2\\
\hline
\end{tabular} 
\end{center}
\label{tab:cap}
\end{table}

From these values, the charge loss to backplane has been 
estimated to be $\simeq$~50\% for the geometries adopted in the present
test structures, thus allowing an effective signal interpolation. 

The charge collection properties of the test sensors have been directly studied
by shining an infrared diode spot on the backplane of a structure with 
60~$\mu$m implant width, 100~$\mu$m implant pitch and 200~$\mu$m read-out 
pitch. 
At the diode wavelength of $\lambda$ = 880~nm, the penetration depth in the 
silicon substrate corresponds to 10~$\mu$m. The IR light has been focused to a
spot size of $\simeq$~80~$\mu$m and its position parallel to the detector 
plane controlled by a 2-D stage allowing to scan the pixel array with 
micro-metric accuracy parallel and orthogonal to the pixel rows. 
A matrix of 4 $\times$ 7 
read-out pixels has been wire-bonded to a VA-1 chip. For each spot position, 
1000 events have been recorded. 
The common mode, pedestal and noise calculation has been initialised for the 
first 300 events. In the subsequent events light was injected every 10 events,
allowing for continuous pedestal tracking. 
Because of the limited data volume, no on-line suppression has been applied 
and the data reduction and cluster search has been performed off-line. 
Results have been averaged over the 70 recorded light pulses. 
Figure~1 shows the recorded pulse height as a function of the scan position. 
The charge loss has been measured by comparing the cluster pulse height for
the laser spot positioned on either an interleaved or a readout pixel. The
results are shown in Figure~1, where  pulse heights are normalised to their
maximum value. The observed charge loss does not exceed 50~\%, in agreement 
with the expectations from the capacitance measurements.

\begin{figure}[hb!]
\begin{center}
\epsfig{file=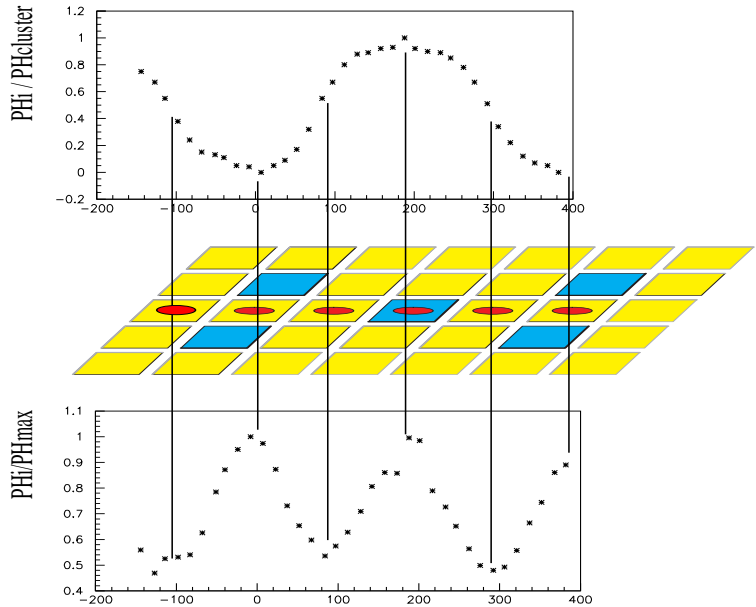,width=10.0cm,height=8.0cm,clip}
\caption{\sl The cluster pulse height, in units of the largest measured value, 
(lower plot) and the charge sharing among neighbouring read-out pixels, 
$\eta = \frac{PH_i}{PH_{tot}}$, (upper plot) measured during the detector 
scan.}
\end{center}
\label{fig:scan}
\end{figure}

The charge sharing can be characterised by the shape of the $\eta$ distribution
defined as $\eta = \frac{PH_i}{PH_{tot}}$, where $PH_i$ is the pulse height 
recorded on the i$^{th}$ read-out pixel in the cluster and $PH_{tot}$ 
the total cluster pulse height. The measured curve, shown in Figure~1, 
exhibits the expected correlation between the distance of the point of
charge creation from the leftmost strip and the $\eta$ response, where 
$\eta = 1$ corresponds to the particle hitting the i$^{th}$ pixel and 
$\eta = 0.5$ to the particle traversing the sensitive volume half-way between
the i$^{th}$ and the (i+1)$^{th}$ read-out pixel. The observed functional 
dependence can be interpreted as the superposition of the effects of the 
charge diffusion and the capacitive charge sharing.

\vspace{-0.25cm} 

\section{Conclusions}

\vspace{-0.5cm} 

Prototype pixel detectors with interleaved pixel cells, aimed at improving 
their single point resolution to match the requirements for applications at 
the future linear collider,
have been designed and manufactured. The results of their electrostatic 
characterisation and the preliminary charge collection studies have confirmed
the validity of this detector concept.

\end{document}